**Vanadium substitution: a simple and efficient way to improve UV sensing in ZnO**

Tulika Srivastava[1], Gaurav Bajpai[1], Gyanendra Rathore[1], Sajal Biring[2], Somaditya Sen[1]*

[1]Metallurgy Engineering and Materials Science, Indian Institute of Technology Indore, Khandwa Road, Simrol, Indore, India

[2]Electronic Engineering Department and Organic Electronic Research Centre, Ming Chi University of Technology, New Taipei City, Taiwan

**Email ids: *** sens@iiti.ac.in

**Abstract:**

UV sensing in pure ZnO is due to oxygen adsorption/desorption process from ZnO surface. Vanadium doping improves UV sensitivity of ZnO. Enhancement in UV sensitivity in doped ZnO is attributed to trapping and de-trapping of electrons at $V^{4+}$ & $V^{5+}$-related defect states. An extra electron in the $V^{4+}$ state is excited under UV illumination while in absence of the same a trapping happens at the $V^{5+}$ state. An insight to the mechanism is obtained by an analytic study of the response phenomenon.

**Introduction:**

Ultraviolet detection is becoming important nowadays related to various important aspects of science/technology associated with health, environment and even space research[1,2]. Sensitive silicon based UV detectors are already available in market. But these detectors require costly visible light filters as they are sensitive to visible light. Faster, more sensitive, cost-effective UV detection is therefore an important research area. GaN, SiC and diamond are promising candidates[3–5]. But all of these are expensive materials. ZnO is an abundant, inexpensive, non-toxic and environmental friendly material with good thermal/chemical stability and high photoconductivity. UV sensing and response in ZnO, mainly depend on the surface reaction and therefore, surface defects, grain size and oxygen adsorption properties[6–8]. Several morphological studies of ZnO show enhancement in UV sensing. Doping on the other hand modifies electronic, optoelectronic and photoconductive properties of ZnO. Significant modifications have been observed in optoelectronic properties with various types of doping. Vanadium doping is one of the most interesting ones exhibiting luminescence, opto-electronic and photo sensing properties. These properties arise out of electron trapping defect states formation within the bandgap. This study analyses the effect of vanadium doping on UV sensing properties of ZnO and establish a probable trap-state mediated mechanism of UV sensing in vanadium doped ZnO system.

**Experimental:**

Vanadium doped ZnO nanoparticles ($Zn_{(1-x)}V_xO$) for x=0 (ZV0), 0.0078 (ZV1), 0.015 (ZV2), and 0.023 (ZV3) have been synthesized by the sol-gel method (standard Pechini method) which is followed by solid state calcination. First of all, ZnO powder was dissolved in $HNO_3$ (Alfa Aesar, purity 99.9%) and dopant $V_2O_5$ was made soluble by dissolving it into $NH_4OH$. Both the solution was then added together and stirred for some time for homogeneous mixing. Citric acid and glycerol were used as a gelling agent in this process. These two polymerize by releasing $H_2O$ from OH groups (esterification) of citric acid and glycerol when heated at 70°C for 1 h [9]. The polymeric solution was added to the ZnO/V solution. The resultant solution was stirred vigorously while being continuously heated at 70°C. Zn and V ions get attached homogeneously to the polymer solution. Evaporation of water in the solution resulted in a gel formation in about 4 h. The gels were burnt on the hot plate in ambient conditions. The resultant powders were decarbonized and denitrified at 450°C for 6 h. These powders were then pressed (3 Tons) into pellets of 0.1mm thickness and 13 mm diameter. The prepared pellets were further annealed at 500°C for 2 h. In all the samples two electrodes were prepared on the surface of pellets using silver paste at a distance of 1.25 mm.

Structural characterization was carried out using x-ray diffractometer (Bruker D2-Phaser). A home-made setup was fabricated to estimate UV light sensing of $Zn_{(1-x)}V_xO$ pellets.

Two conductive stainless-steel wires were used as connectors from the electrodes to the positive and negative terminal of Keithley (2401) meter as shown in figure 2(a). To take care that no stray light can affect the experiment the sample setup was kept inside a dark box. A steady UV LED light source of 390 nm wavelength was focussed from the top of the black box. The sample to source distance was maintained at 10cm such that the intensity at the surface of the sample was always ~270 lux. Five cycles of UV ON/OFF sequences of 7.5 min each were studied by recording the dynamic changes in current.

**Result and Discussion:**

XRD spectra revealed a dominant hexagonal wurtzite ZnO structure with some minor reflections similar to zinc blend as shown in Figure 1(a). No secondary phases were found related to simple or complex oxides of Zn and V for all samples (with vanadium doping from

x=0 to x =0.023) heat treated at 500˚C. Higher doping (above ZV3) is not reported due to formation of $Zn_2V_2O_7$ phases in minor quantities.

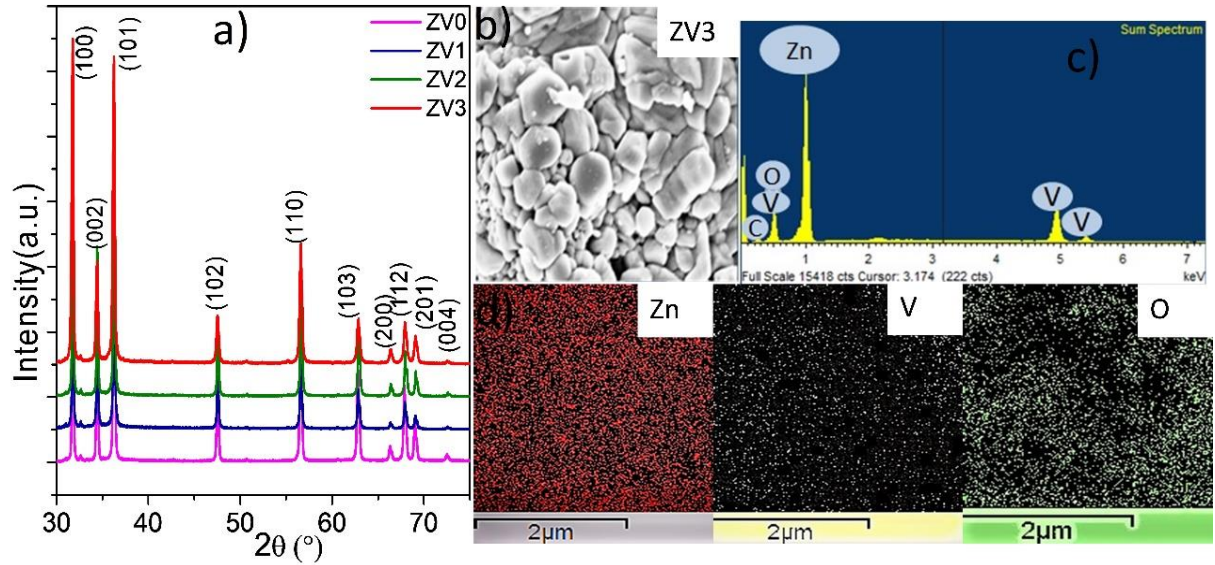

Figure 1: (a) XRD spectra of ZV0, ZV1, ZV2 and ZV3 (b) SEM image of ZV3 (c) EDS of ZV3 (d) Elemental mapping of ZV3

FESEM and EDS results [Figure 1 (b, c, d)] show the compact topology and elemental analysis of a certain area of ZV3 sample. A 2D elemental scan of the sample over the same area reveals uniform distribution of Zn, O and V. This trend is just not for a single area of a specific sample but is a feature of all regions in all the samples. This indicates that vanadium has been uniformly doped substituting Zn in the ZnO lattice. The atomic percentage of Vanadium for ZV1, ZV2 and ZV3 was found to be 0.71%, 1.41% and 1.76% which is quite similar to our doping percentage.

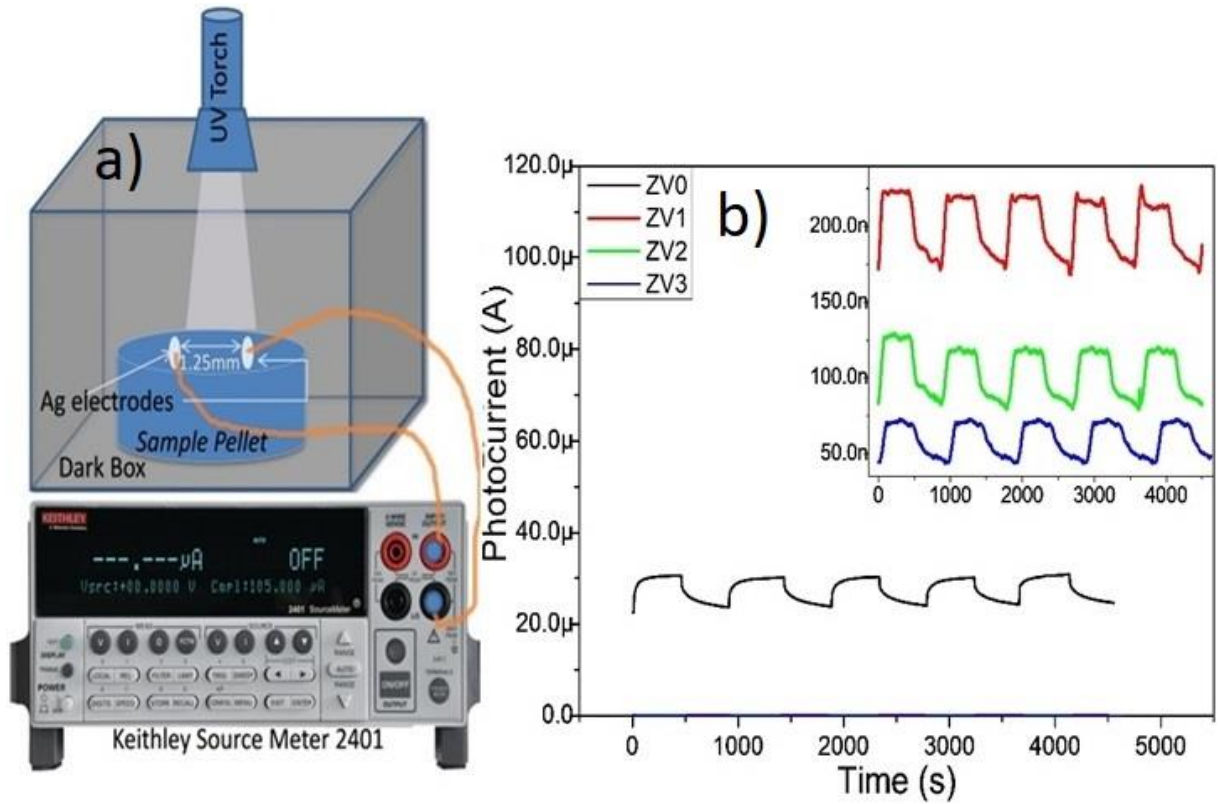

Figure 2: (a) Home-made set up of UV sensing (b) Photocurrent response of ZV0, ZV1, ZV2 & ZV3.

Figure 2 (b) shows the photo response of current of undoped and vanadium doped ZnO when measured at a bias voltage of 8V. In the "ON" state, i.e. with UV light falling on the sample, there is significant increase in current, $I_{UV}$, in comparison to the current in "OFF" state, $I_{Dark}$, when the sample is in darkness. The sensitivity in current when subject to UV light can be used for sensing similar wavelengths. Figure shows five UV "ON/OFF" cycles. The cycles are similar in nature and repeatable. This confirms the reliability of sensitivity of the synthesized materials.

In pure ZnO in absence of light the dark current is due to the electron-hole pairs generated due to the external voltage (8V) applied. In case of doped ZnO the electrons in the conduction band gets easily trapped by new defects states created by interactions of V3d and O2p orbitals. This reduce electron transport and thereby conductivity of the material[10] .Hence, it was anticipated that there will be decrease in the dark current with vanadium doping. This is exactly what is observed in these ZVO samples. However, although dark current was lesser in vanadium incorporated samples, with UV illumination, the percentage change in current increased.

To estimate such properties, UV sensitivity was defined by the formula[11,12]:
$$S = \frac{I_{UV} - I_{dark}}{I_{dark}} \times 100$$
Sensitivity was plotted for all samples [Figure 3 (a)] and was found to increase with increasing vanadium content.

Surface oxygen defects play an important role in UV-sensitivity. It has been discussed in general, that adsorption and desorption of oxygen molecules from the surface of a material, is affected by UV radiation[13]. In dark condition, oxygen molecules capture free electrons from the surface of pellets and get adsorbed [$O_2(g) + e^- \rightarrow O_2^-$ (ad)]. This leads to formation of high resistive region near the surface of the pellets. When the surface is illuminated by UV light of energy (390 nm) corresponding to bandgap of material, electron-hole pairs are generated [$h\nu \rightarrow e^- + h^+$]. Photo generated holes react with adsorbed oxygen, thereby liberating the oxygen from the surface [$h^+ + O_2^-$ (ad) $\rightarrow O_2(g)$] making the electron available for surface transport. Therefore, availability of oxygen defect sites plays an important role on the free carrier concentration facilitating a larger current upon UV illumination. However, in vanadium doped ZnO samples due to the higher valence state of vanadium, oxygen will be retained in the lattice more than pure ZnO. As a result oxygen vacancies, $V_0^{\bullet\bullet}$, should reduce and zinc replaced $V^{5+ \text{ or } 4+}$ sites, $V_{zn}^{\bullet\bullet\bullet \text{ or } \bullet\bullet}$, should increase [Kroeger and Vink notation[14]]: $2V^{5+/4+} + 5/4 O^{2-} + 5/4 V_0^{\bullet\bullet} \rightarrow 2V_{zn}^{\bullet\bullet\bullet/\bullet\bullet} + 5/4 O_o^X$; where, $O_o^X$ is an oxygen ion in its site. Oxygen content was found to increase in the substituted samples, as revealed from XANES studies [15]. Hence, if surface oxygen defects would have been the mechanism then with reduced oxygen vacancies, sensitivity should have reduced with increased vanadium incorporation. But in contradiction, there is enhancement in sensitivity. This clearly signifies that oxygen adsorption and desorption is not the main reason for sensitivity in these samples.

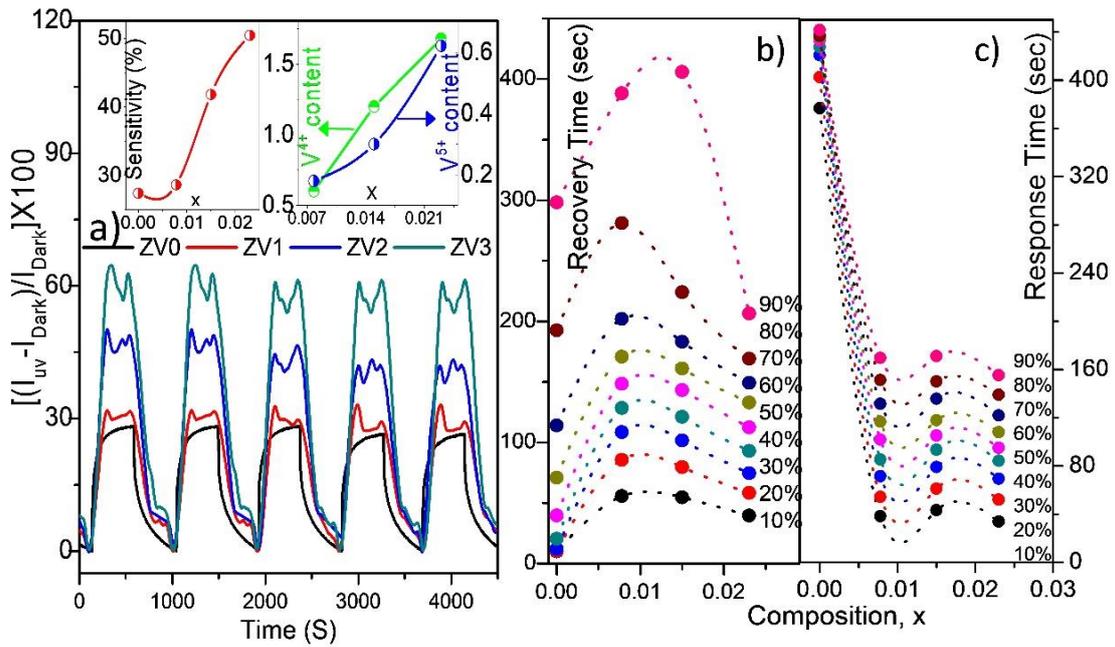

Figure 3: (a) Sensitivity of ZV0, ZV1, ZV2 & ZV3 [inset is $V^{4+}$ and $V^{5+}$ content with x] (b) Variation of Recovery Time with x (c) Variation of Response Time with x.

We have estimated the overall $V^{4+}$ and $V^{5+}$ content in all the samples using the formula, $V^{4+\ or\ 5+}$ content = Composition (x). Fraction of $V^{4+or5+}$. XANES analysis provided an estimation of ratio of $V^{4+}$ and $V^{5+}$ content[15]. Both $V^{4+}$ and $V^{5+}$ contents show increasing trends similar to sensitivity as shown in inset of figure 3 (a). To be noted that the trend of increment in sensitivity is strikingly similar to that of $V^{4+}$ ions. Hence vanadium ($V^{4+}$ or $V^{5+}$) incorporation must be creating defect states in between the valence and conduction bands of ZnO[15–17], which trap free electrons from the lattice thereby reducing the dark current. While $V^{5+}$ state is empty, $V^{4+}$ state has one available electron. When illuminated, these $V^{4+}$ electrons transit to the conduction band and contribute to the photocurrent. On the other hand, after removal of UV illumination, the excited electrons require empty $V^{5+}$ states to transit and get trapped [$V^{5+} + e^{-} \rightarrow V^{4+}$]. Trapping is easier with abundance of empty $V^{5+}$ states. Note that with such a trapping process the current decreases and gradually approach dark current as dark current is a result of transport of available carriers in the conduction band only after trapping is saturated.

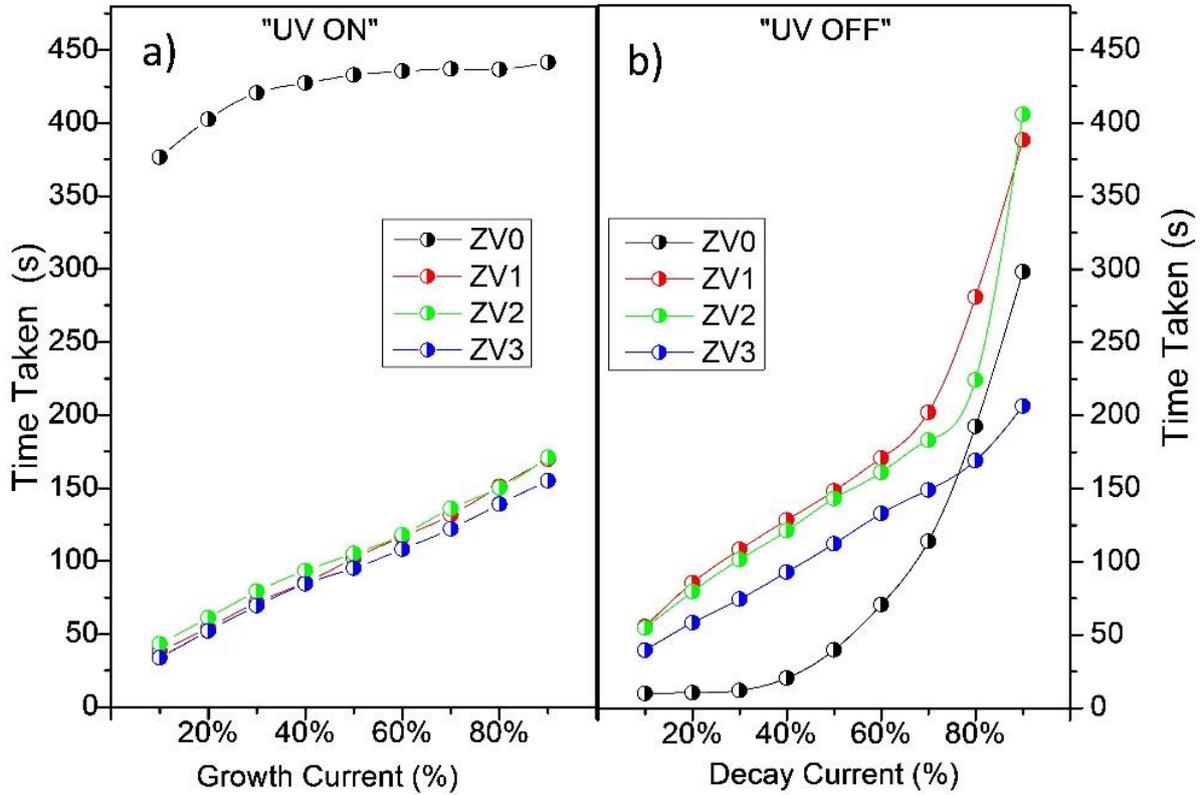

Figure 4: (a) Variation in response time for each percentage of current growth (b) Variation in recovery time for each percentage of current growth

The trapping phenomenon is dependent on how fast electrons can be trapped by empty defect states in the bandgap. Response time (time required to reach 10%, 20%, 30%, 40%, 50%, 60%, 70%, 80% and 90% of final value of current in the presence of UV light) and recovery time (time required to reduce the current to 10%, 20%, 30%, 40%, 50%, 60%, 70%, 80% and 90% of saturation value in the absence of UV light) were calculated for each sample shown in Figure 3 (b, c). Note that the time taken to return to the basic dark current from a saturated photo-excited current is larger in case of doped samples than the pure ZnO samples. With higher substitution the rate becomes faster indicating the availability of more defect states for trapping. The photo-response is faster in vanadium doped samples than pure ZnO. From this observation it may be inferred that de-trapping of electrons from the V-related defect states to the conduction band in V-doped ZnO is faster than adsorption/desorption rate of oxygen in ZV0. Note that the response is fastest in the ZV3 sample as compared to ZV1 and ZV2, implying that availability of more $V^{4+}$ electrons actually contribute to promptness of photocurrent generation.

Response time was slower for pure ZnO (ZV0) for all percentages [Figure 4 (a, b)]. Recovery time was faster for the same than the doped samples. Note that the trend in percentage growth graphs is linear for doped samples in contrast to pure ZnO samples, clearly hinting at different mechanisms for UV sensing in these samples. Oxygen desorption and de-trapping of V d-shell electrons, in pure and doped ZnO respectively are thereby probable mechanisms behind photo response, with the latter being faster.

**Conclusion:**

Sol gel prepared $Zn_{(1-x)}V_xO$ samples, for $0<x<0.023$, show enhanced UV sensitivity with vanadium doping. Difference between the mechanism of sensing in between the pure and doped samples has been discussed. In doped ZnO trapping and de-trapping of electrons by V-related defects in ZnO lattice is proposed as a possible mechanism, while in pure ZnO it is mostly due to surface oxygen defect related mechanism. Photosensitivity improves with higher doping. Trends in response and recovery time with composition help develop a new method in analysis of the mechanism. Vanadium doping is a simple, economic and efficient way to improve UV sensing property of ZnO.

**Acknowledgement:**

The authors thank SIC (IIT Indore) for providing FESEM facility.

**References:**

[1] E. Monroy, F. Omnès, F. Calle, Wide-bandgap semiconductor ultraviolet photodetectors, Semicond. Sci. Technol. 18 (2003) R33. doi:10.1088/0268-1242/18/4/201.
[2] K. Liu, M. Sakurai, M. Liao, M. Aono, Giant Improvement of the Performance of ZnO Nanowire Photodetectors by Au Nanoparticles, J. Phys. Chem. C. 114 (2010) 19835–19839. doi:10.1021/jp108320j.
[3] L. Sang, M. Liao, M. Sumiya, A Comprehensive Review of Semiconductor Ultraviolet Photodetectors: From Thin Film to One-Dimensional Nanostructures, Sensors. 13 (2013) 10482–10518. doi:10.3390/s130810482.
[4] T. Toda, M. Hata, Y. Nomura, Y. Ueda, M. Sawada, M. Shono, Operation at 700°C of 6H-SiC UV Sensor Fabricated Using N+ Implantation, Jpn. J. Appl. Phys. 43 (2003) L27. doi:10.1143/JJAP.43.L27.
[5] E. Muñoz, (Al,In,Ga)N-based photodetectors. Some materials issues, Phys. Status Solidi B. 244 (2007) 2859–2877. doi:10.1002/pssb.200675618.
[6] A.J. Gimenez, J.M. Yáñez-Limón, J.M. Seminario, ZnO−Paper Based Photoconductive UV Sensor, J. Phys. Chem. C. 115 (2011) 282–287. doi:10.1021/jp107812w.
[7] S. Hong, J. Yeo, W. Manorotkul, G. Kim, J. Kwon, K. An, S.H. Ko, Low-Temperature Rapid Fabrication of ZnO Nanowire UV Sensor Array by Laser-Induced Local Hydrothermal Growth, J. Nanomater. (2013). doi:10.1155/2013/246328.

[8] S. Bai, W. Wu, Y. Qin, N. Cui, D.J. Bayerl, X. Wang, High-Performance Integrated ZnO Nanowire UV Sensors on Rigid and Flexible Substrates, Adv. Funct. Mater. 21 (2011) 4464–4469. doi:10.1002/adfm.201101319.

[9] A.E. Danks, S.R. Hall, Z. Schnepp, The evolution of "sol–gel" chemistry as a technique for materials synthesis, Mater. Horiz. 3 (2016) 91–112. doi:10.1039/C5MH00260E.

[10] M.K. Gupta, J.-H. Lee, K.Y. Lee, S.-W. Kim, Two-Dimensional Vanadium-Doped ZnO Nanosheet-Based Flexible Direct Current Nanogenerator, ACS Nano. 7 (2013) 8932–8939. doi:10.1021/nn403428m.

[11] H. Zhang, Y. Hu, Z. Wang, Z. Fang, L.-M. Peng, Performance Boosting of Flexible ZnO UV Sensors with Rational Designed Absorbing Antireflection Layer and Humectant Encapsulation, ACS Appl. Mater. Interfaces. 8 (2016) 381–389. doi:10.1021/acsami.5b09093.

[12] W. Dai, X. Pan, C. Chen, S. Chen, W. Chen, H. Zhang, Z. Ye, Enhanced UV detection performance using a Cu-doped ZnO nanorod array film, RSC Adv. 4 (2014) 31969–31972. doi:10.1039/C4RA04249B.

[13] S.P. Ghosh, K.C. Das, N. Tripathy, G. Bose, D.H. Kim, T.I. Lee, J.M. Myoung, J.P. Kar, Ultraviolet photodetection characteristics of Zinc oxide thin films and nanostructures, IOP Conf. Ser. Mater. Sci. Eng. 115 (2016) 012035. doi:10.1088/1757-899X/115/1/012035.

[14] F.A. Kröger, H.J. Vink, Relations between the Concentrations of Imperfections in Crystalline Solids, Solid State Phys. 3 (1956) 307–435. doi:10.1016/S0081-1947(08)60135-6.

[15] T. Srivastava, G. Bajpai, N. Tiwari, D. Bhattacharya, S.N. Jha, S. Kumar, S. Biring, S. Sen, Opto-electronic properties of Zn(1-x)VxO: Green emission enhancement due to V4+ state, J. Appl. Phys. 122 (2017) 025106. doi:10.1063/1.4992087.

[16] J. Guo, W. Zhou, P. Xing, P. Yu, Q. Song, P. Wu, Structural, magnetic and optical properties of vanadium doped zinc oxide: Systematic first-principles investigations, Solid State Commun. 152 (2012) 924–928. doi:10.1016/j.ssc.2012.03.016.

[17] E. García-Hemme, K. M. Yu, P. Wahnon, G. González-Díaz, W. Walukiewicz, Effects of the d-donor level of vanadium on the properties of Zn1−xVxO films, Appl. Phys. Lett. 106 (2015) 182101. doi:10.1063/1.4919791.